\documentstyle[12pt,aaspp4]{article}
\slugcomment{Submitted to {\it The Astrophysical Journal}}
\lefthead{Morales  et al.}
\righthead{FUV Spectra of B stars}

\begin{document}

\title
{Far Ultraviolet Spectra of B Stars near the Ecliptic\footnote{Based on the 
development and utilization of the Espectr\'ografo
Ultravioleta de Radiaci\'on Difusa, a collaboration of the Spanish Instituto 
Nacional de T\'{e}cnica Aeroespacial and the Center for EUV Astrophysics, 
University of California, Berkeley}}
\author{Carmen Morales, Ver\'{o}nica Orozco, Jos\'{e} F. G\'{o}mez} 
\affil{Laboratorio de Astrof\'{\i}sica Espacial y F\'{\i}sica
  Fundamental, INTA, 
  Apdo. Correos 50727, E-28080 Madrid, Spain}
\author{Joaqu\'{\i}n Trapero}
\affil{Universidad SEK, Cardenal 
Z\'{u}\~{n}iga s/n, E-40003 Segovia, Spain, and Laboratorio de 
Astrof\'{\i}sica Espacial y F\'{\i}sica
  Fundamental, INTA}
\author{Antonio Talavera}
\affil{Laboratorio de Astrof\'{\i}sica Espacial y F\'{\i}sica
  Fundamental, INTA}
\author{Stuart Bowyer, Jerry Edelstein, Eric Korpela, Michael Lampton}
\affil{Space Sciences Laboratory, University of California,
Berkeley, CA 94720--7304}
\author{Jeremy J. Drake}
\affil{Harvard-Smithsonian Center for Astrophysics,
MS-3, 60 Garden Street, Cambridge, MA 02138}

\begin{abstract}

Spectra of B stars in the wavelength range of 911-1100 \AA\ have been obtained
with the
EURD spectrograph onboard the Spanish satellite MINISAT-01 with {$\sim$}
5 \AA\ spectral resolution. IUE spectra of the same stars have been used to normalize
Kurucz models to the distance, reddening and spectral type of the corresponding star.
The comparison of 8  main-sequence stars studied in detail   
($\alpha$ Vir, $\epsilon$ Tau, $\lambda$ Tau, $\tau$ Tau, $\alpha$ Leo, $\zeta$ Lib, 
$\theta$ Oph, and $\sigma$ Sgr) shows agreement with Kurucz
models, but observed fluxes are 10-40\% higher than the models 
in most cases. The difference in
flux between observations and models is higher in the wavelength range between
Lyman $\alpha$ and Lyman $\beta$. We suggest that Kurucz
models underestimate the FUV flux of main-sequence B stars
between these two Lyman lines. Computation of flux distributions of
line-blanketed model atmospheres including non-LTE effects suggests that this
flux underestimate could be due to departures from LTE, although other causes 
cannot be ruled out.
We found the common assumption of solar
metallicity for young disk stars should be made with care, since small
deviations can have a significant impact on FUV model fluxes.
Two 
peculiar stars ($\rho$ Leo and $\epsilon$ Aqr), 
and  two emission line
stars ($\epsilon$ Cap and $\pi$ Aqr) were also studied.  Of these, 
only $\epsilon$ Aqr 
has a flux in agreement with the models. The rest 
have strong variability in the
IUE range and/or uncertain reddening, which makes the 
comparison with models difficult.

\end{abstract}


\section{Introduction}

The far ultraviolet (FUV) spectral region is one of the least studied
ranges in stellar physics. 
IUE and HST
have provided a large number of stellar spectra at wavelengths longer
than 1100 \AA. However, data at wavelengths below 1100 \AA\ are 
much more limited.
Observations by Voyager (Broadfoot at al. 1977\markcite{bro77}), 
Copernicus (Rogerson
et al. 1973\markcite{rog73}), HUT (Davidsen 1990\markcite{dav90}), ORFEUS 
(Hurwitz \& Bowyer 1991\markcite{hur91}),
and UVSTAR (Stalio et al. 1993\markcite{sta93}), and rocket observations
by Brune, Mount,
\& Feldman (1979\markcite{bru79}), Carruthers, Heckathorn, \& Opal 
(1981)\markcite{car81}, have provided some data.
The absolute fluxes obtained by these investigations are often highly
discrepant for a variety of reasons.

In a previous work (Morales et al. 2000\markcite{mor00}), 
we reported on FUV
spectroscopic observations of the bright star $\alpha$ Vir (Spica)
obtained with the EURD spectrograph, 
which provided an absolute flux for this star.
EURD data on $\alpha$ Vir were reasonably well reproduced
by Kurucz models (Kurucz 1993\markcite{kur93}) fitted to IUE data.

In this paper, we present spectra of stars observed by EURD
during its first two years of operation. Although the design of this
spectrograph is optimized to study diffuse radiation, it can detect
bright, early-type stars in the FUV. 
In \S\ref{observations} we present the observations,
data reduction, and our use of IUE data. 
In 
%\S\ref{main} 
\S 3
we present our results for main sequence stars
and compare them with model atmospheres. 
In \S\ref{peculiar} 
%\S 4
we discuss the peculiar and emission line stars observed.
We summarize our conclusions in \S\ref{conclu}.
%\S 5.

\section{Observations, Data Reduction and Merged IUE Spectra}
\label{observations}
EURD is one of the instruments onboard the Spanish satellite
MINISAT-01. It consists of two spectrographs covering the bandpass
between 350 and 1100 {\AA}, with {$\sim$} 5 {\AA} spectral resolution. 
MINISAT-01 was launched on April 21, 1997. 
Its orbit is retrograde with an inclination of 151$^\circ$ 
and an altitude of 517 km. 
The design of the spectrographs and their
ground calibrations are described in detail in Bowyer, Edelstein, \& Lampton
(1997\markcite{bow97}). 

EURD samples the entire ecliptic plane once per year
using nightly observations in the anti-sun direction.
The EURD spectrographs perform simultaneous observations 
with a field of view of 25$^\circ \times$  8$^\circ$. 
Stars with ecliptic latitudes between --13$^\circ$
and +13$^\circ$ can, in principle, be detected by EURD.
However, only those between --4$^\circ$ and +4$^\circ$ 
are certain to fall into the field of view. 
The 12 stars presented in this paper are those
which have been 
observed up to March 31, 1999.
Other stars were detected in these EURD data, however
obtaining their spectra was impossible 
because of spectral confusion from other stars 
simultaneously in the field of view. EURD has detected stellar emission
only from 912 to 1100 {\AA} due to the absorption by the interstellar medium.

The data reduction process consists of identifying and separating stellar 
point-source emission from the instrumental noise and diffuse emission 
produced by strong geocoronal airglow emission lines and possibly
by the hot component of the interstellar medium.
The photon counting detector records images with spectral
resolution ($\sim$ 5 {\AA}) in one axis and angular resolution 
($\sim 10 \arcmin$) in the other axis. 
When a star falls in the field of view of the instrument, 
its spectra is distributed within a finite area in the angular dimension. 
In contrast, instrumental background and diffuse emission will be
located along the entire angular dimension.
To test for the presence of a stellar source, 
we accumulated an entire detector image every 28 seconds.
The detector image was integrated over the spectral dimension
to obtain a histogram  of total spectral intensity as a function 
of field-view angle.
If the maximum of the histogram exceeded 4 times the noise level,
defined as the standard deviation of the continuum level of the histogram, 
then a stellar spectrum was obtained from the flux
found within 7 angular pixels ($10\farcm5$) of the maximum.
A net stellar spectrum was derived
by subtracting the atmospheric and instrumental background 
taken from an equivalent area adjacent to the stellar region.
The net spectrum's intensity was corrected using
the inflight-calibrated variation 
of optical efficiency as a function of radiation input angle 
(determined with observations of $\alpha$ Vir during April 1999).
Intensity-corrected spectra were co-added for each stellar object
and converted to fluxes, applying the
in-flight calibration performed 
using simultaneous observations of the full Moon
with EUVE and EURD. 
Further details of the data reduction process can be found
in G\'omez et al. (2000\markcite{gom00}).

\begin{figure}
\epsscale{0.8}
\plotone{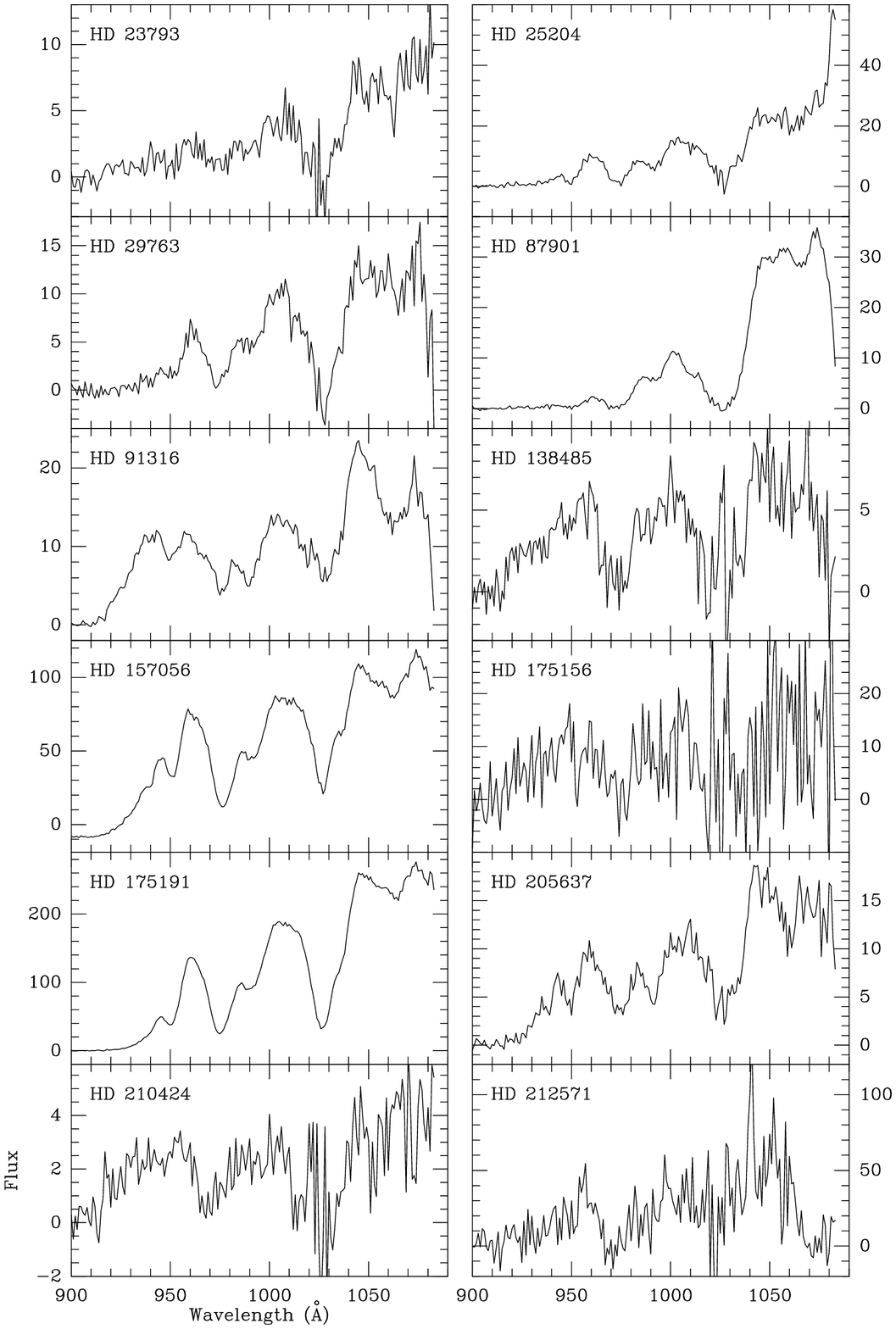}\caption{\label{figspectra} Spectra of the stars observed
by EURD. The y axis represents the flux in units of $10^{-10}$ erg
s$^{-1}$ cm$^{-2}$ \AA$^{-1}$.}
\end{figure}

\label{results}
We have obtained spectra of 12 B stars taken over two years of
observations.
Table \ref{tabdates}
shows the dates of the observations. The
total integration times and statistical significance are shown  
in Table \ref{tabnoise}. 
%The IUE spectra used are in Table 3.
Fig. \ref{figspectra} shows the spectra for
all the stars except HD 116658 ($\alpha$ Vir), 
which has been previously published in 
Morales et al. (2000\markcite{mor00}).

We have divided the stars into three groups:
main-sequence, peculiar, and emission line
stars. The parameters of the stars in these groups are shown in Table
\ref{tabstars}. Spectral types and 
V magnitudes are from the Simbad database, color excesses are 
obtained using the intrinsic (Warren 1976\markcite{war76}) 
and observed (b-y) color 
(Hauck \& Mermilliod 1998\markcite{hau98}), and the relation of Penprase 
(1992\markcite{pen92}) between E(b-y) and E(B-V). 
Effective temperatures are deduced from the comparison of Kurucz 
models (see \S 3) 
%\ref{kuruczuse} 
only for main-sequence stars, due to the uncertainties 
found in model fitting for peculiar and
emission line stars (see \S \ref{peculiar} below).

IUE spectra of the observed stars were selected from the INES database, according 
to their quality. Only large 
aperture, not trailed observations were taken. In the case of HD 157056 the only 
available short
wavelength IUE spectrum is of small aperture and has been normalized to fit the large
aperture long wavelength spectrum of this star.
Spectra from SWP and LWP cameras were joined at 1940 \AA, truncating SWP spectra
longward of 1940 \AA\ and LWP spectra shortward of 1940 \AA. IUE
spectra used in this paper are shown in Table \ref{tabIUEspec}.
The IUE absolute flux calibration defined for the final archive used in this work 
is 7.2 -- 10 \% lower than the Hubble Space 
Telescope flux scale (Gonz\'alez-Riestra 1998\markcite{gon98}, Massa \& Fitzpatrick
2000\markcite{mas00}).

\section{Main Sequence Spectra and Comparisons with Model Atmospheres}
\label{compar}

We have compared our data with the Kurucz (1993\markcite{kur93}) atmosphere models, 
extracted from the homogeneous grid of models of Lejeune, Cuisinier, \&
Buser (1997\markcite{lej97}). The comparison with Kurucz models is done with 
the same method we used for $\alpha$ Vir, binning IUE and EURD spectra to the 10 \AA\ 
spectral resolution of the models and dereddening them with the 
Cardelli, Clayton, \& Mathis (1989\markcite{car89}) extinction law. 
Between 912 and 1100 {\AA} this law was extrapolated by polynomial fit, as
suggested by Longo et al. (1989\markcite{lon89}). However instead of
fixing the temperature, which is not well known for the stars of the present sample,  
we took the temperature of the model that best fits the IUE range longward of 
1246 {\AA} (in order to avoid the Lyman $\alpha$ geocoronal component present in 
IUE spectra), using the IDL program KSCALE of the IUE RDAF Library, that searches for
the minimum rms deviation between models and data. Thus, the temperatures derived by 
this fit do not depend on absolute flux, only on the shape of the spectra.
The values of the stellar 
radii deduced from the fit to IUE (assuming Hipparcos distances, ESA 
1997\markcite{esa97}) are within the values,
obtained by different methods, reported by Fracassini et al. (1988\markcite{fra88}). 

Atmosphere models of solar composition, microturbulent velocity $v_{t}$ = 2.0,
and gravity log $g$ = 4.0 were
taken for main-sequence 
stars. A ratio of total to selective
extinction R = 3.1 was used for all the stars except for HD 138485 
which is located in
the Upper Scorpius association. Guti\'errez-Moreno \& Moreno 
(1968\markcite{gut68}) found for this part of the association a value of R = 6. 
Using Hipparcos distances (ESA 1997\markcite{esa97}) and more recent spectral 
types from the Simbad database for 44 stars belonging to the Upper Scorpius 
association, we also found R=6 for this region.
This value was used to deredden HD 138485.

\begin{figure}
\epsscale{0.8}
\plotone{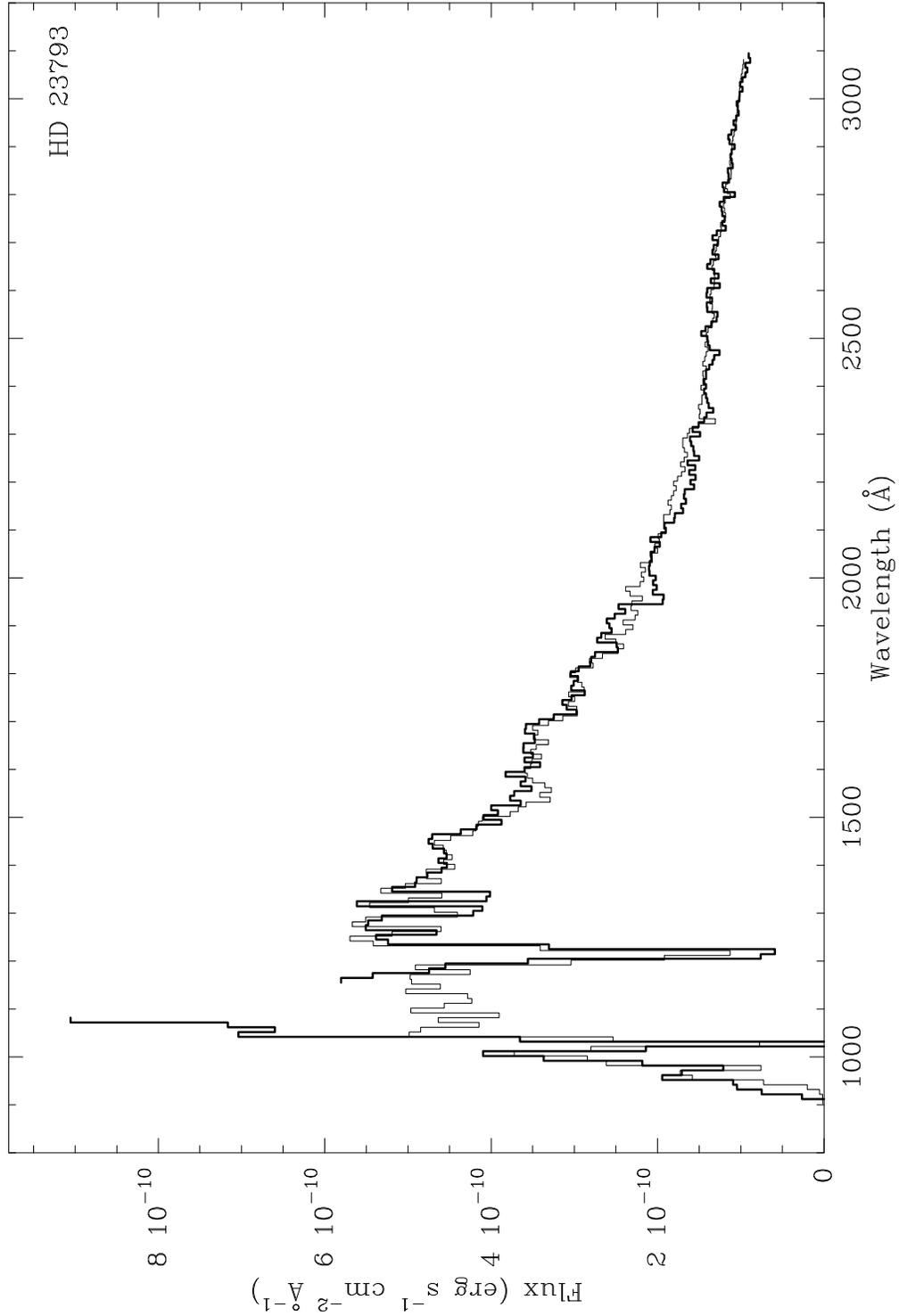}\caption{\label{figiuemodel} Spectra of HD 23793 as
observed by EURD ($\lambda <  1070$ \AA) and IUE ($\lambda > 1160$ \AA)
in heavy lines, 
superposed on the Kurucz model (thin line) that best fit the IUE data. Both 
EURD and IUE spectra have been binned down to match 
the spectral resolution of the model (10 \AA).} 
\end{figure}

\begin{figure}
\epsscale{0.8}
\plotone{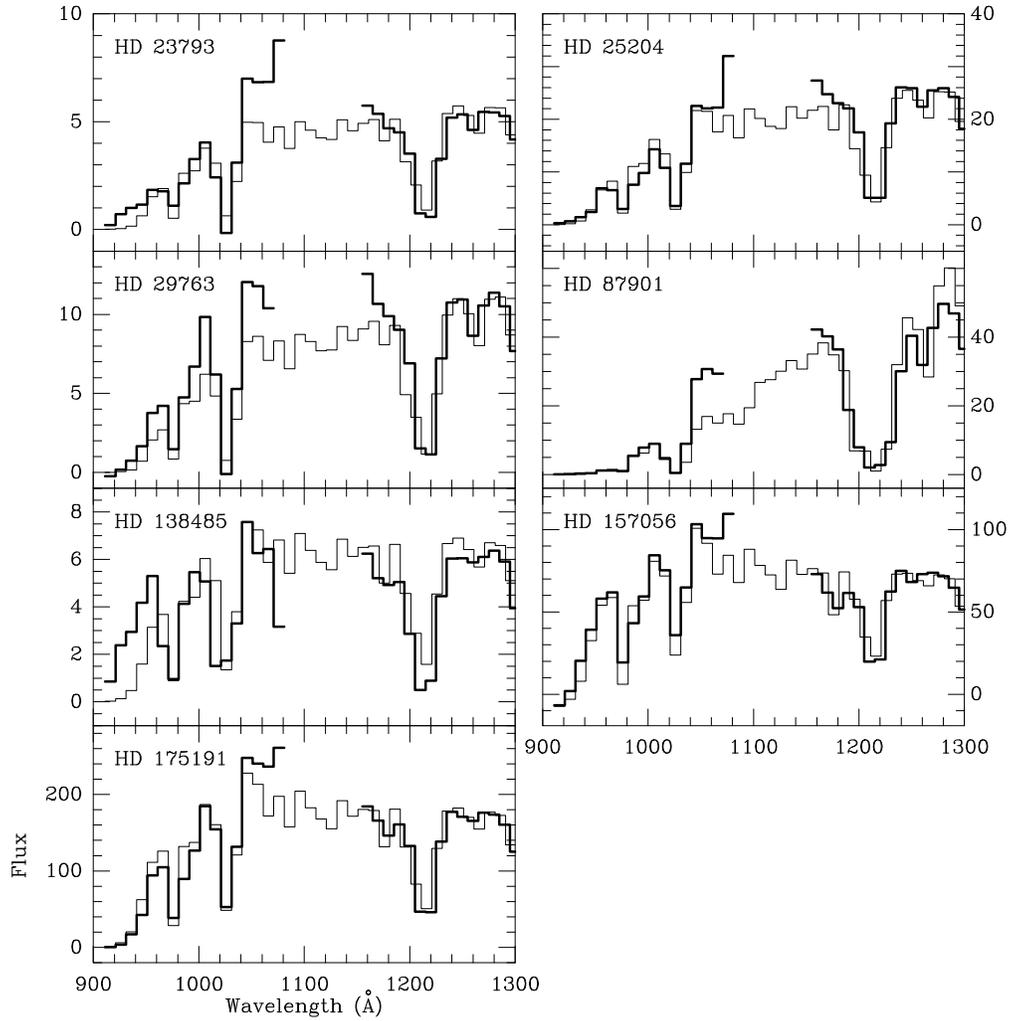}\caption{\label{figeurdmodel} Spectra of main-sequence stars 
as observed by EURD (heavy line) and the model (thin line) that best fit the 
IUE data (heavy line, $\lambda > 1100$ \AA). 
The y axis represents the flux in units of $10^{-10}$ erg
s$^{-1}$ cm$^{-2}$ \AA$^{-1}$.}
\end{figure}

In Fig. \ref{figiuemodel} we show the reddened Kurucz model that best fits
the IUE range, and the whole IUE and EURD spectra for HD 23793. 
Fig. \ref{figeurdmodel} shows the comparison of EURD and IUE data with
models for all main-sequence stars.
EURD spectra and the appropriate Kurucz models show a general agreement, 
with the best agreement 
occurring at wavelengths shorter 
than Lyman $\beta$ (1026 \AA). EURD integrated fluxes shortward of
1010 \AA\ are 10 to 40\% higher than 
Kurucz fluxes in most cases (see Table 3), while for wavelengths 
longward of Lyman $\beta$, the EURD flux can exceed the model values 
by up to 95\% for the worst case (HD 87901 [$\alpha$ Leo], see Fig. \ref{figeurdmodel}). 
This confirms the result obtained for EURD 
observations of $\alpha$ Vir (Morales et al. 2000\markcite{mor00}) that 
also found higher fluxes than expected from the models.
If we correct the 
IUE absolute flux calibration to match the HST flux calibration, the
agreement with the models improves except for the stars with negative
difference (HD 25204 and HD 175191). 
Only HD 138485 shows slightly less flux than the 
models longward of Lyman $\beta$, but this could reflect the
uncertainty in the value of of R=6 used for dereddening.

\begin{figure}
\epsscale{0.8}
\plotone{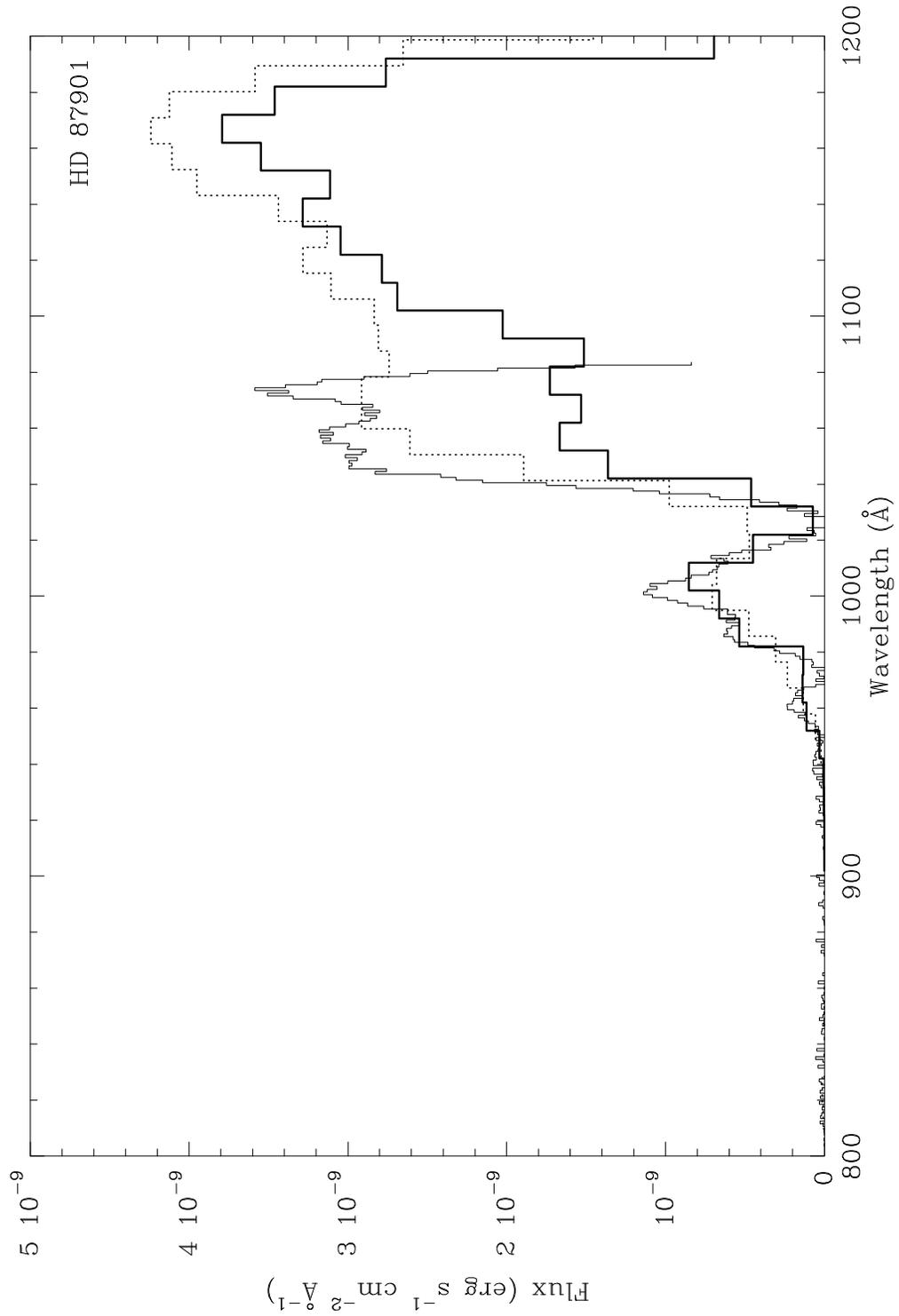}\caption{\label{fighd87} Spectra of  HD 87901 ($\alpha$
Leo) as observed
by EURD (thin line, $\lambda <  1070$ \AA) (thin line) and Voyager (dotted line), 
superimposed on the Kurucz model (heavy line) that best fit the IUE spectrum.}
\end{figure}

The same trend of observed fluxes higher than the models can be seen in IUE 
spectra shortward of Lyman $\alpha$ (1216 \AA) for HD 23793, HD 25204 and HD
29763, and HD 87901, although high fluxes at longest EURD wavelengths and 
shortest IUE ones could be due to edge effects in the detectors. In fact, 
errors have been found to be systematically higher at IUE shortest wavelengths 
(Massa \& Fitzpatrick 2000\markcite{mas00}). However, Voyager data, which 
have the region between Lyman $\alpha$ and Lyman $\beta$ in the middle 
of their detectors, agree with EURD fluxes for the two stars they have 
in common, $\alpha$ Vir and $\alpha$ Leo.  
Fig. \ref{fighd87}  shows a
comparison of EURD and Voyager observations with the corresponding
model for $\alpha$ Leo, which shows the highest discrepancy from models 
at the longest EURD wavelengths. It is obvious that both observations
closely agree with each other, and are significantly higher than the
model. The agreement between Voyager and EURD is also remarkable 
for $\alpha$ Vir 
(Morales et
al. 2000\markcite{mor00}), 
both being higher than the corresponding
atmosphere model.

Main-sequence B stars are thought to be relatively
well understood in terms of the modeling of their photospheres.
However, as discussed earlier, there have been relatively 
few studies in the FUV to test available models.  
The discrepancies we found between observed and model fluxes could in
principle arise from several different sources: EURD instrument
calibration errors; inappropriate ISM reddening corrections;
deficiencies in the Kurucz LTE model atmosphere; or errors in
the adopted model stellar parameters.  

An incorrect EURD calibration might be responsible for the discrepancy.  
However, the detailed character of our 
inflight calibration and the agreement in observed fluxes for 
two stars, $\alpha$~Vir and $\alpha$~Leo, observed by both {\it Voyager} 
and EURD, using independent calibrations (with white dwarfs and the Moon 
respectively), give support to the correctness of our calibration.

Significant problems associated with our extinction corrections 
also seem unlikely.  While accurate corrections for the effects of ISM
absorption and scattering in the FUV is a difficult problem 
(e.g., Savage et al. 1985\markcite{sav85}),
all of our stars have very low values of reddening and consequently any
uncertainty will be small.  
Though early results from a
survey of interstellar H$_2$ made by FUSE 
have indicated an ubiquitous presence of H$_2$
(Tumlinson et al. 1999\markcite{tum99}),
we have not applied
corrections for H$_2$ absorption because the H$_2$ columns toward
these targets are small and the effects should be negligible.
We have verified this assumption by computing the H$_2$ absorption in the
900-1100~\AA\ range seen at the resolution of the EURD instrument.
The relatively narrow molecular bands begin to saturate at quite low
H$_2$ columns ($<10^{18}$~cm$^{-2}$).  However, they account for
relatively little net flux loss at these columns when integrated over
the EURD response function.  This same result was found by Snow, Allen \& Polidan
 (1990\markcite{sno90}) for the case of {\it Voyager} spectra: for a H$_2$ column
of $10^{18}$~cm$^{-2}$ (a value higher than that expected for any of
the main-sequence stars in our sample) the net absorption amounts to
less than 5~\% . We do note that
a value of R=1.0 would increase model FUV fluxes without distorting
the IUE range very much, but this value of R is very unlikely (He et al. 
1995\markcite{he95}) 

We suggest that Kurucz atmosphere models underestimate fluxes
between Lyman $\alpha$ and Lyman $\beta$. 
It is worth noting that other authors studying stellar spectra in 
this wavelength range have also found discrepancies in the same direction 
between their observations and Kurucz models (e.g., Chavez, Stalio, \& Holberg 
1995\markcite{cha95}; Buss, Kruk, \& Ferguson 1995\markcite{bus95};
Dixon \& Hurwitz 1998\markcite{dix98}). 
 
An important assumption used in the Kurucz model
atmospheres adopted in this study 
is that of LTE.  In order to investigate the effects
of relaxing this assumption, 
we have carried out new calculations 
using the TLUSTY model atmosphere program 
(Hubeny \& Lanz
1995\markcite{hub95}; Hubeny 2000, private communication) 
that uses extensive line-blanketing and that 
treats H, He and
abundant metals fully in non-LTE in both line and continuum processes.
TLUSTY model atmospheres for representative effective temperatures of
15000~K and 20000~K, with a surface gravity of $\log g=4.0$ and
solar metallicity ([M/H]$=0.0$), were computed both under the assumption
of LTE and also with the LTE assumption relaxed.  In addition, a slightly 
cooler model with effective temperature of 13500~K was added because
of the rapid change in this temperature regime 
in important continuum opacity sources from
neutral C, N and O to once ionized species in the FUV region.

\begin{figure}
\epsscale{0.8}
\plotone{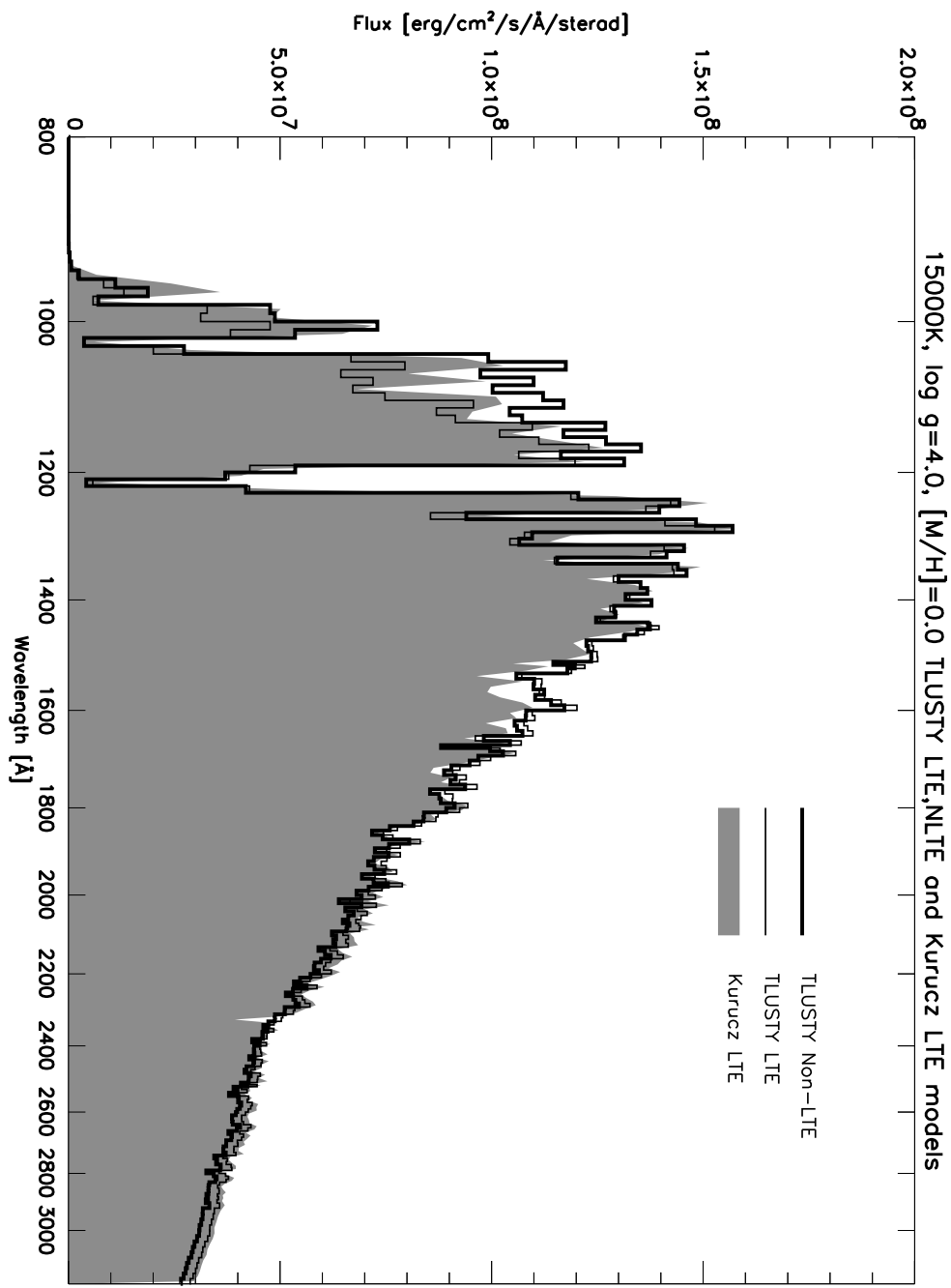}\caption{\label{f:lte_nlte_kur_15} Comparison between
 synthetic spectra computed
using TLUSTY LTE and non-LTE models for parameters $T_{eff}=15000$, 
$\log g=4.0$, $[M/H]=0.0$.  The corresponding Kurucz model
flux is illustrated by the shaded region.}
\end{figure}

In terms of temperature structure, the LTE and non-LTE models are very
similar in the deepest layers at all the effective temperatures
investigated, as expected.  Toward higher layers in the line-forming
regions, the LTE models are slightly hotter than the non-LTE models by
up to 400~K or so.  In the outer layers this situation is reversed,
with the non-LTE models ending up hotter but by a larger amount of
about 1500~K.  We have computed the resulting synthetic spectra using
the LTE and non-LTE models; comparisons for the 15000~K models are
illustrated in Figure~\ref{f:lte_nlte_kur_15}, binned at a
resolution commensurate to that of the IUE spectra used to determined
the stellar effective temperatures.  We also illustrate the
corresponding Kurucz model as a grey shaded region.

For the IUE spectral range, although the forms of the LTE and non-LTE
model spectra appear quite similar, many differences in the various
more prominent absorption lines become readily apparent when examining
these model spectra at high spectral resolution.  At the lower
resolution of the observations however, these differences become
washed out and the emergent fluxes in each spectral bin are in general similar.
The situation changes in the FUV range of interest
for our EURD observations: it is immediately apparent that the non-LTE
model fluxes are substantially higher overall than the LTE fluxes,
especially in the region between Lyman $\alpha$ and $\beta$.  For the
FUV spectral range between Lyman $\alpha$ and $\beta$, differences
between non-LTE and LTE model fluxes increase from longer to shorter
wavelengths, amounting to as much as 30~\%\ or so near 1050~\AA .
Toward wavelengths shortward of Lyman~$\beta$, LTE and
non-LTE model fluxes converge slightly and are in somewhat better
agreement.

Comparisons between hotter and cooler temperature
models reveal an interesting trend.  In the case of the hotter 20000~K
model, the differences between LTE and non-LTE fluxes are in the same
sense as for the 15000~K model, but they are significantly smaller and amount
to only 10~\%\ or so near 1050~\AA .  Agreement throughout the IUE
range is also slightly better.  In contrast, comparison of the
slightly cooler 13500~K models reveals {\it even larger differences}
between LTE and non-LTE fluxes than for the 15000~K models, amounting
to approximately a factor of 2 at 1050~\AA .  In all cases the
differences are reduced again toward wavelengths shortward of
Lyman~$\beta$.  We note that the trend of increasing disparity between
LTE and non-LTE model fluxes with decreasing effective temperature,
especially near 1050~\AA , is similar to the discrepancies between
observations and LTE models, which are most pronounced for the coolest
star of the sample.

We also note that the wavelength regime in which the non-LTE
effects appear most significant supports our method of deriving
effective temperatures based on IUE fluxes longward of Lyman~$\alpha$,
where non-LTE effects appear comparatively small at IUE resolution.
At the same time, our findings suggest difficulties for methods hoping
to use FUV fluxes for the derivation of effective temperatures based
on synthetic LTE spectral indices.

Making the straightforward conclusion that non-LTE effects must be
responsible for the model and observed FUV flux disparities in our
program main-sequence B stars is complicated by the comparison between
the TLUSTY LTE and Kurucz LTE models.  It is apparent from
Figure~\ref{f:lte_nlte_kur_15} that these are also not in good
agreement in the FUV range. The Kurucz model predicts more FUV
flux than the TLUSTY model.  The reason for these discrepancies is not
obvious, though it may be related to differences in the line
and continuum opacities employed.  Taking just the LTE models, the
Kurucz models are in better agreement with the observations than
the TLUSTY LTE models.  However the 
{\it differential} comparison of TLUSTY LTE and non-LTE models is valid and
these non-LTE effects will be in operation in late B star
photospheres.  Therefore, while the problem as to the best model
formulation for the FUV range must remain unanswered, we conclude that
at least some, and perhaps all,
of the differences we are observing between model and
EURD fluxes are caused by departures from LTE in the photosphere.

The particular departures from LTE that give rise to the effects
described above are in the continua of low Z metals that dominate the
continuous opacity in the FUV region in late B stars.  Neutral species such
as C{\sc i} are overionized in the photospheric layers, and so the opacity
is reduced leading to higher fluxes in the regions affected.  Toward
higher effective temperatures the neutrals become more minor species
and cease to dominate the continuous opacity; departures from LTE in
these species are therefore less important.

We have also investigated the sensitivity of the Kurucz
model FUV fluxes to the stellar parameters: effective temperature,
surface gravity and metallicity.  There are surprisingly few modern,
detailed, high resolution spectroscopic studies available for 
main-sequence B field stars on which to draw for guidance as to the exact
parameters and compositions of our target stars.  As has been
discussed by previous authors, the effective temperature scale of B
main-sequence stars is dependent to a significant extent on the
indices used to derive it, even down to different UV indices (e.g.
Chavez et al. 1995\markcite{cha95}; Buss et al. 1995\markcite{bus95}).
Moreover, it is well known
that the FUV fluxes of B main-sequence stars are very sensitive to the
effective temperature adopted.  Indeed, the 500-1000~K higher
effective temperatures derived by Buss et al. (1995\markcite{bus95}) 
based on {\it HUT}
FUV fluxes are a direct result of the same excess observed flux at
short wavelengths as is evident in our EURD spectra.  What is less
often discussed is the effects of surface gravity and metallicity on
these fluxes (see Fitzpatrick \& Massa 1999\markcite{fit99}, for a discussion on
their effects on UV-optical fluxes).

\begin{figure}
\epsscale{0.8}
\plotone{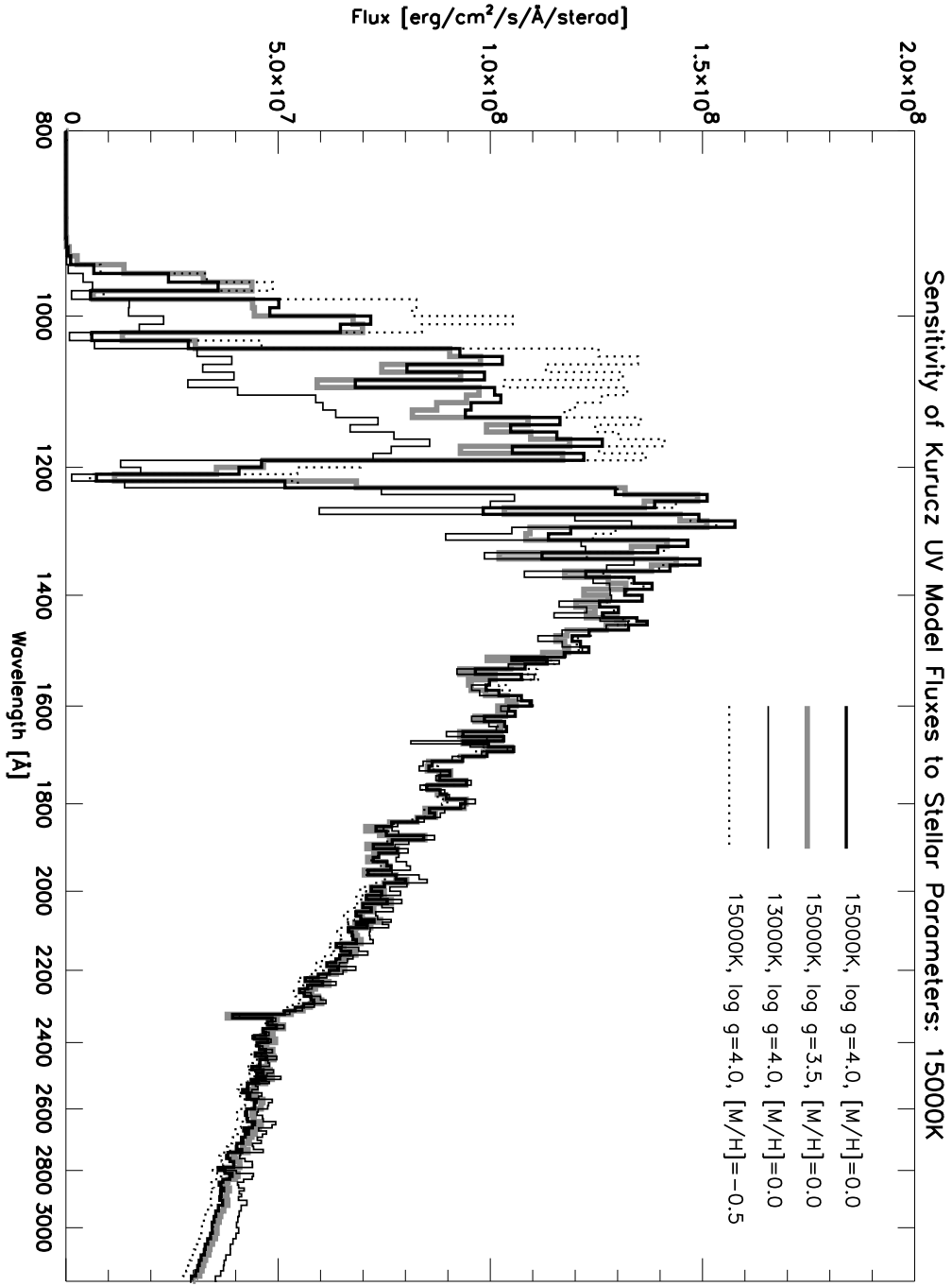}\caption{\label{f:mod15_plots} Comparison between Kurucz 
model UV and FUV
fluxes to illustrate their sensitivities to stellar parameters.
Illustrated are parameters 15000~K, $\log g=4.0$, [M/H]$=0.0$, and
perturbations to this set orthogonally in temperature ($-2000$~K),
surface gravity ($-0.5$) and metallicity ($-0.5$).}
\end{figure}

We illustrate the FUV flux sensitivity to model parameters for
the case of the $T_{eff}=15000$~K model in Figure~\ref{f:mod15_plots}. 
As expected, strong sensitivity to temperature is shown by the
comparison with a model 2000~K cooler, but with otherwise similar
parameters.  The flux disparity increases toward shorter wavelengths
as expected from the analogous black body behavior.  
A difference of 2000~K in effective temperature is much larger
than any uncertainties in the derivation of temperature scales, but it
is clear that even adopting model parameters a few hundred degrees
different would significantly alter the predicted fluxes in the 
very shortest wavelengths. The flux discrepancy we are seeing
for $\alpha$~Leo is not simply an error in the adopted
effective temperature. This is clear in view of the nature of the
discrepancy: instead of increasing with decreasing wavelength, the
model and observations are actually in much better agreement below
1000~\AA\ than above.

It is also clear from Figure~\ref{f:mod15_plots} that stellar FUV
fluxes are not strongly dependent on the exact value of surface
gravity adopted.  Reducing the surface gravity leads to a {\em
decrease} in the FUV flux and a slight {\em increase} in the flux at
the longest UV wavelengths.  Thus, for the case of a typical B star in
our study, an error in the adopted surface gravity is likely to be at
least partially compensated for by the derivation of a slightly
different effective temperature. The resulting predicted FUV fluxes
{\em seen at the resolution of our instrument} would be corrected for
by a compensatory temperature error.

The value adopted for the metallicity is much more important.
Previous FUV studies appear to have assumed that the metallicities of all
early-type stars are identically equal to the solar value 
(Chavez et al. 1995\markcite{cha95}; Buss et al. 1995\markcite{bus95}).  Studies of
disk stars and young stars formed in very similar ISM environments
show this not to be the case at levels that could be significant for
FUV fluxes.  The studies of abundances in main-sequence B stars in the
Orion association by Cunha \& Lambert (1992\markcite{cun92}, 
1994\markcite{cun94}) revealed
differences in O and Si abundances of as much as 0.2 dex in stars
co-located on the sky and at similar distances.  They conjectured that
these differences arose because they were formed from different
regions of a parent molecular cloud enriched to different extents with
the ejecta of Type II supernovae.  On a much wider scale, young stars
in the solar neighborhood also exhibit significant scatter in
metallicity amounting to $\pm 0.2$~dex or so about the solar value
(e.g. Luck \& Lambert 1985\markcite{lac85}; Nissen
1988\markcite{nis88}; Boesgaard 1989\markcite{boe89}; 
Edvardsson
et al. 1993\markcite{edv93}).  
Assumption of solar metallicity for young disk stars
should therefore only be done under peril of errors of up to 50 \%.  

By inspection of Figure~\ref{f:mod15_plots}, we note a difference in
FUV flux shortward of Lyman~$\beta$ of as much as 30~\% for a model
differing in metallicity by 0.5~dex, with the change occurring in the
obvious sense that lower metallicity, leading to lower FUV opacity,
leads to higher FUV fluxes.  Consequently, care should be taken in the
adoption of the metallicity parameter if comparison is to be
meaningful at a level of order 15~\% or better.

Another possible stellar parameter that can have an important effect on 
FUV fluxes is microturbulent velocity. Fitzpatrick \& Massa 
(1999\markcite{fit99}) have found a dependence of UV flux on this parameter 
that is of the same order as the effect of surface gravity and metallicity. 
However, their calculation does not extend to the FUV. Therefore it would 
be interesting to investigate the impact of microturbulent velocity in the 
EURD range. 

In the particular case of $\alpha$ Leo, we note that its rotational velocity 
(v = 350 km $s^{-1}$, Bernacca \& Perinotto 1970) could also be important to 
interpret its large FUV flux excess. Model calculations by Collins \& Sonneborn 
(1977\markcite{col77}) predict a flux dependency on rotational velocity that 
becomes more significant in the FUV, for B0-F8 stars. However this effect has 
not been found observationally (Molnar, Stephens \& Mallama 1978\markcite{mol78},
Llorente de Andr\'es \& Morales 1979\markcite{llo79}).
On the other hand, it has been suggested that rotational velocities may induce an
effective color excess for spectral types earlier than B6 ( Maeder
1975\markcite{mae75}, Llorente de Andr\'es et al. 1981\markcite{llo81}), which
if neglected, would result in a spurious ultraviolet excess. This would not be the
case for $\alpha$ Leo, a B7 star, and we suggest that its high rotational velocity do
not have a significant impact on its FUV flux excess.

\section{Peculiar Stars and Emission Line Stars }
\label{peculiar}
We obtained spectra of two peculiar stars.
However, the strong influence of metallicity in 
the far ultraviolet fluxes of model
atmospheres makes comparisons with models difficult for these
stars. An accurate knowledge of this stellar parameter is necessary in
order to find the appropriate model to compare with the observations.
The two peculiar stars we observed were:

{\bf HD 210424}. 
A suspected chemically peculiar star of the Si type in the catalog of Renson,
Gerbaldi, \& Catalano (1991\markcite{ren91}). 
Cayrel de Strobel et al. (1997\markcite{cay97}) found a metal abundance of 
[Fe/H] $= -0.26$, while Leone, Manfr\'e, \& Catalano (1995\markcite{leo95}), 
with optical high resolution spectroscopy, could not find any sign of 
metallicity, while the overall flux
distribution could be reproduced
using a Kurucz model (ATLAS9) with solar abundance. This, together with flux
differences of 20\% among IUE observations at $\lambda > 1400$ \AA\ makes the
comparison with model atmospheres difficult. With both values of metallicity 
EURD fluxes are two to three times higher than any possible model fitted to the IUE 
wavelength range. 

{\bf HD 91316}. An OBN supergiant with moderated nitrogen enhanced (Walborn
1976\markcite{wal76}),
with a metal abundance of [Fe/H] $= -0.89$ (Cayrel de Strobel et al.
1997\markcite{cay97}). After 
interstellar reddening correction of E(B-V) $=0.056$, 
the EURD fluxes are well fit by a Kurucz
model of 19400 K, log g=3.0 and with a metallicity of -0.89.

\label{emline}
We obtained spectra of two emission line stars. In the study of these stars,
the difficulty in evaluating the amount of emission from the central star and
from the circumstellar envelope makes the determination of the 
interstellar extinction very complicated. Both
stars in our sample show
strong variability in the IUE observations. Since these observations were
not simultaneous with our EURD observations, any comparisons
are problematical.
The two emission line stars observed were:

{\bf HD 205637}. 
An emission line star, also classified as peculiar of the Si type by
Renson et al.
(1991\markcite{ren91}).
Neither its metal abundance, nor its real color excess (it has significant 
circumstellar envelope emission) are known. A well-known shell star, 
MWC 373 (Merrill \& Burwell 1933\markcite{mer33}),
it is a very close system with components of magnitudes 4.9 and 6.2 at less 
than 0\farcs01 
plus another component of 9.5 at $68''$ (Catalano \& Renson 1998\markcite{cat98}). 
It is a variable star with a 
period of 0.9775 days (Pedersen 1979\markcite{ped79}). IUE observations show a flux 
variability of a factor of two. 

{\bf HD 212571}. Classified as a
shell star, MWC 388 (Merrill \& Burwell 1933\markcite{mer33}), 
IUE observations from 1979 to 1995 show a variability of 
$\sim$ 38 \% for the whole short-wavelength IUE range. 
Assuming that its energy distribution is unaffected by envelope 
radiation in the 3700-5500 \AA\ spectral region, Kaiser (1989\markcite{kai89})
deduced an E(B-V) = 0.11 for the central star. Kurucz models fitted to maximum and
minimum IUE data show a difference of 600 K between them, and the extension of none of
them fit well the EURD range. EURD fluxes are higher than the two models fitted and
also brighter than Voyager and HUT (Buss et al. 1994\markcite{bus94}) observations of
this star.

\section{Conclusions}
\label{conclu}
We have obtained the far ultraviolet spectra (shortward of 1100 \AA)
of 12 B-type stars, using the EURD spectrograph on-board MINISAT-01.
We have carried out a detailed comparison of these spectra with several
model atmosphere codes.  We find:

\begin{enumerate}

\item Comparison with Kurucz models show that EURD fluxes for main-sequence B
stars are 10-40 $\%$ higher than the model predictions in most cases 
for wavelengths shorter than Lyman $\beta$. This difference would be reduced 
using the HST calibration to correct IUE fluxes.

\item  For wavelengths longer than Lyman $\beta$ EURD spectra tend to 
show a stronger flux
excess compared with Kurucz model atmospheres for both IUE and HST 
absolute flux scales. A flux excess between Lyman $\alpha$ and $\beta$
is also seen in IUE spectra for half of our main-sequence stars.
%as well as in the Voyager spectra of two stars of our sample they
%have observed. We suggest that Kurucz models underestimate
%the FUV flux of main-sequence B stars, especially between  
%Lyman $\alpha$ and $\beta$.
%Previous FUV observations: Voyager, HUT and ORFEUS have reported also an 
%excess flux in its observations compared to Kurucz model predictions.

\item Comparisons between flux distributions of line-blanketed model
atmospheres 
indicate that the non-LTE case is overionized relative to the LTE case.
Non-LTE effects on low Z metals, which are important continuum opacity sources,
can lead to elevated fluxes shortward of the C{\sc i} edge.  
We therefore suggest
that the observed flux excesses compared to Kurucz LTE models between
Lyman~$\alpha$ and $\beta$ could be due to departures from LTE.  
The non-LTE effects increase with decreasing effective 
temperature as the neutral absorbers,
though significantly overionized, become major species.
At higher effective temperatures these absorbers are completely 
ionized and do not contribute to the opacity.
\item The common assumption of solar metallicity for young disk
stars should be questioned when attempting to model the FUV flux of
mid- and late-type B stars.  Even fairly small deviations from the solar
mixture of 0.1-0.2~dex (values typical of the scatter expected in 
present-day star-forming regions and in young stars in the solar 
neighborhood) can have a significant and observable
impact on FUV fluxes that are heavily moderated by metal line
blanketing.
\item For peculiar and emission line stars a precise determination of variability and
metallicity is necessary to perform good comparisons between models and observations.
In our data of these stars only HD 91316 could be successfully fit. 
\end{enumerate}

\acknowledgments

We are grateful to Jay Holberg for sending us the Voyager data.
This research has made use of the SIMBAD database, operated at CDS, Strasbourg,
France, and of the International Ultraviolet Explorer data retrieved 
from the INES Archive.
The development of EURD has been partially supported by 
INTA grant IGE490056. V.O. and J.T. acknowledge support by
Junta de Castilla y Le\'{o}n, grant SEK1/00B. JFG is supported
in part by DGESIC grant PB980670-C02 and by Junta de Andaluc\'{\i}a
(Spain). 

Partial
support for the development of the EURD instrument was provided 
by NASA grant NGR05-003-450. When NASA funds were withdrawn 
by Ed Weiler, the instrument was completed
with funds provided by S. Bowyer. The UCB analysis and 
interpretation in this work was carried
out through the volunteer efforts of the authors.
JJD would like to extend warm thanks to Ivan Hubeny for help with 
computation of the non-LTE TLUSTY model atmospheres.
JJD was supported by the Chandra X-ray Center NASA contract NAS8-39073
during the course of this research.

\clearpage

\clearpage

\begin{deluxetable}{lccl}
\tablecaption{\label{tabdates} Observation log}
\tablehead{
   \colhead{HD number} & \colhead{Year} & \colhead{Month} & \colhead{Days}
}
\startdata
HD 23793 &  1997 & November & 10--12 
\nl
HD 25204 &  1997 & November & 11--13 
\nl
HD 29763 &  1997 & December & 5--7 
\nl
HD 87901 &  1998 & February & 18-- 2 
\nl
        & 1999 & February & 16, 19--20 
\nl
HD 91316 &  1998 & February & 23--25 
\nl
        & 1998 & March & 1--4, 6--8 
\nl
        & 1999 & February & 27 
\nl
        & 1999 & March & 7--8 
\nl
HD 116658 & 1998 & April & 13--19
\nl
          & 1999 & April & 2--23 
\nl
HD 138485 & 1998 & May & 11, 13--18 
\nl
HD 157056 & 1998 & June & 13--17 
\nl
HD 175156 & 1998 & June & 28--30 
\nl
HD 175191 & 1998 & July & 3--7, 9, 11, 30   
\nl
HD 205637 & 1997 & August & 14--17, 21--22
\nl
         & 1998 & August & 19 
\nl
HD 210424 & 1998 & August & 23--27, 29 
\nl
HD 212571 & 1997 & August & 23
\nl
\enddata

\end{deluxetable}

\begin{deluxetable}{lccr}
\tablecaption{\label{tabnoise} Observational parameters of the spectra}
\tablehead{
   \colhead{HD number} & \colhead{Total integration time} & \colhead{ noise
   level $\sigma$}\tablenotemark{a}\tablenotetext{a}{Measured between 750 and 850
\AA. Valid for wavelengths $\la 945$ \AA. For longer wavelenghts, this noise
level increases to a factor of 2 at $\sim 1070$ \AA.} & 
\colhead{signal to noise}\tablenotemark{b}\tablenotetext{b}{Signal measured at 1000
\AA .}\\
        & \colhead{(10$^4$s)} & \colhead{(10$^{-11}$erg s $^{-1}$
   cm $^{-2}$ \AA $^{-1}$ )}&
}
\startdata
HD 23793 & 1.1150 & 6.2 & 7
\nl
HD 25204 & 2.1128 & 4.9 & 31 
\nl
HD 29763 & 1.6725 & 4.9 & 20
\nl
HD 87901 & 7.3273 & 2.4 & 45 
\nl
HD 91316 & 8.3627 & 2.9 & 44 
\nl
HD 116658& 10.9550& 2.6& 3662
\nl
HD 138485 & 1.7477& 8.0 & 11
\nl
HD 157056 & 4.3473& 3.7 & 219
\nl
HD 175156 & 0.3353& 39.7 & 3 
\nl
HD 175191 & 5.5884& 3.2 & 514 
\nl
HD 205637 & 4.6017& 4.1 & 29 
\nl
HD 210424 & 3.0884& 5.1 & 8 
\nl
HD 212571 & 0.4358& 64.5 & 6
\nl
\enddata

\end{deluxetable}

\begin{deluxetable}{lccccccc}
\tablecaption{\label{tabstars} Stellar parameters}
\tablehead{\colhead{HD number} & \colhead{Name} & \colhead{Spectral 
   Type} & \colhead{V} & \colhead{E(B-V)} & \colhead{$T_{\rm eff}$}
\tablenotemark{a}\tablenotetext{a}{For peculiar and
emission line stars no appropriate model could be
fit.} &
   \colhead{EURD -- Kurucz}\tablenotemark{b}\tablenotetext{b}{Relative difference
 between EURD and Kurucz model for $\lambda<1010$ \AA\ and $\lambda>1040$
([EURD -- Kurucz]/Kurucz).} \\
        &&&&& (K) & $\lambda<1010$ & $\lambda>1040$\\
}
\startdata
\cutinhead{Main Sequence Stars}
HD 23793 & $\epsilon$ Tau & B3 V & 5.06 & 0.0449 & 17800 & 0.24 & 0.57
\nl
HD 25204 & $\lambda$ Tau & B3 V & 3.47 & 0.0582 & 18100 & -0.11 & 0.21
\nl
HD 29763 & $\tau$ Tau & B3 V & 4.29 & 0.0389 & 16600 & 0.425 & 0.54
\nl
HD 87901 & $\alpha$ Leo & B7 V & 1.33 & 0.0382 & 13200 & 0.10 & 0.95
\nl
HD 138485 & $\zeta$ Lib & B3 V & 5.50 & 0.0453 & 19100 & 0.36 & -0.15
\nl
HD 157056 & $\theta$ Oph & B2 IV & 3.26 & 0.0114 & 21700 & 0.08 & 0.13
\nl
HD 175191 & $\sigma$ Sgr & B2.5 V & 2.078 & 0.00 & 19900 & -0.13 & 0.22
\nl
\cutinhead{Emission Line Stars}
HD 205637 & $\epsilon$ Cap & B3 Vpe & 4.70 & 0.0087 & \nodata & \nodata 
\nl
HD 212571 & $\pi$ Aqr & B1 Ve & 4.64 & 0.11 & \nodata & \nodata
\nl
\cutinhead{Peculiar stars}
HD 91316 & $\rho$ Leo & B1 Ib & 3.85 & 0.056 & \nodata& \nodata
\nl
HD 175156 &  & B3 II & 5.10 & 0.266 & \nodata & \nodata
\nl
HD 210424 & $\epsilon$ Aqr & B5 III & 5.42 & 0.0329 &\nodata &\nodata
\nl
\enddata
\end{deluxetable}

\begin{deluxetable}{lccr}
\tablecaption{\label{tabIUEspec} IUE spectra from the INES archive}
\tablehead{\colhead{HD number} & \colhead{IUE Short Wavelength spectra} &
\colhead{IUE Long Wavelength spectra} 
}
\startdata
HD 23793  & SWP20583RL.FITS & LWR16502RL.FITS \nl
HD 25204  & SWP18283HL.FITS & - \nl
HD 29763  & SWP45931HL.FITS & - \nl
HD 87901  & SWP33624LL.FITS & LWP08231LL.FITS \nl 
HD 116658 & SWP33091HL.FITS & LWP13650HL.FITS \nl
HD 138485 & SWP35528LL.FITS & LWP15008LL.FITS \nl
HD 157056 & SWP04430RS.FITS & LWP24025RL.FITS \nl
HD 175191 & SWP16368RL.FITS & LWR12623RL.FITS \nl
\nl
\enddata
\end{deluxetable}

\end{document}